\documentclass[aps,
showpacs,
preprint,
superscriptaddress]{revtex4}
\usepackage{amsmath}
\usepackage{epsf}
\begin{document}
\title {Structural Properties of Small Semiconductor-Binding Synthetic Peptides}
\author{G\"{o}khan G\"{o}ko\u{g}lu}
\email[E-mail: ]{Goekhan.Goekoglu@itp.uni-leipzig.de}
\affiliation{Institut f\"ur Theoretische Physik, Universit\"at Leipzig,
Augustusplatz 10/11, D-04109 Leipzig, Germany}
\affiliation{Hacettepe \"Universitesi, Fizik M\"uhendisli\u{g}i B\"ol\"um\"u, Beytepe
06800 Ankara, Turkey}
\affiliation{Centre for Theoretical Sciences (NTZ) of the Centre for Advanced Study (ZHS),
Universit\"at Leipzig, Emil-Fuchs-Stra{\ss}e 1, D-04105 Leipzig, Germany}
\author{Michael Bachmann}
\email[E-mail: ]{Michael.Bachmann@itp.uni-leipzig.de}
\affiliation{Institut f\"ur Theoretische Physik, Universit\"at Leipzig,
Augustusplatz 10/11, D-04109 Leipzig, Germany}
\affiliation{Centre for Theoretical Sciences (NTZ) of the Centre for Advanced Study (ZHS),
Universit\"at Leipzig, Emil-Fuchs-Stra{\ss}e 1, D-04105 Leipzig, Germany}
\author{Tar{\i}k \c{C}elik}
\email[E-mail: ]{tcelik@hacettepe.edu.tr}
\affiliation{Hacettepe \"Universitesi, Fizik M\"uhendisli\u{g}i B\"ol\"um\"u, Beytepe
06800 Ankara, Turkey}
\affiliation{Centre for Theoretical Sciences (NTZ) of the Centre for Advanced Study (ZHS),
Universit\"at Leipzig, Emil-Fuchs-Stra{\ss}e 1, D-04105 Leipzig, Germany}
\author{Wolfhard Janke}
\email[E-mail: ]{Wolfhard.Janke@itp.uni-leipzig.de}
\homepage[\\ Homepage: ]{http://www.physik.uni-leipzig.de/CQT.html}
\affiliation{Institut f\"ur Theoretische Physik, Universit\"at Leipzig,
Augustusplatz 10/11, D-04109 Leipzig, Germany}
\affiliation{Centre for Theoretical Sciences (NTZ) of the Centre for Advanced Study (ZHS),
Universit\"at Leipzig, Emil-Fuchs-Stra{\ss}e 1, D-04105 Leipzig, Germany}
\begin{abstract}
We have performed exhaustive multicanonical Monte Carlo simulations of three 12-residue synthetic
peptides in order to investigate the thermodynamic and structural properties
as well as the characteristic helix-coil transitions. In these studies, we employ a realistic 
model where the interactions between all atoms are taken into account. Effects of solvation are 
also simulated by using an implicit solvent model.
\end{abstract}
\maketitle
\section{Introduction}
\label{intro}
It is well known that three-dimensional (3D) conformations of peptides and proteins play
an important role due to their biological activities. Therefore, many theoretical and
experimental studies focus on the determination of the 3D structure of
these molecules. In a newly growing field of research, synthetic peptides are
investigated for use in hybrid nano-devices, depending on their self-assembly properties~\cite{White,Kamien}.
In these studies, it is also shown that the binding of peptides
on metal and semiconductor surfaces depends on the types of amino acids~\cite{willet1} and on the 
sequences of the residues in the peptide chain~\cite{Brown,Goede,Whaley}. 
These experiments reveal many different interesting and important problems, which are
related to general aspects of the question why and how proteins fold. This regards,
for example, the character of the adsorption process, i.e., whether the peptides simply 
dock to the substrate without noticeable structural changes or whether they perform 
conformational transitions before binding. A related question is how secondary structures 
of peptide folds in the bulk influence the binding behavior to substrates. In helical 
structures, for example, side chains are radially directed and -- due to the helical symmetry --
residues with a certain distance in the sequence arrange linearly. This could have 
consequences for docking to a regular crystal surface, where the atoms are also arranged linearly
along the main axes. This means that two peptides with the same content of residues, but 
different sequences, could exhibit completely different binding properties. This behavior
was, in fact, observed in a recent experimental adsorption study of peptides in the 
vicinity of semiconductor substrates~\cite{Goede}, although other explanations for
this kind of specificity are also conceivable.
For these reasons, it is likely that the binding specificity also depends 
on the thermodynamic and structural properties of peptides in solvent, as it has already turned out 
in investigations of minimalistic models~\cite{MbWj1,MbWj2}.

In this paper, we focus on three synthetic peptides, AQNPSDNNTHTH,
AQNPSDNNTATA, and TNHDHSNAPTNQ~\cite{code}, whose binding properties were
investigated in recent experimental studies~\cite{Goede,Whaley}. The second sequence is a mutated version of
the first one, where the histidine residues (H) of the first chain are
replaced by alanines (A). The third sequence is a randomly permuted sequence of
the first chain. It was shown in the experiments that the first sequence has a strong affinity to adsorb 
to gallium arsenide (GaAs), whereas the binding to silicon (Si) is very weak. 
Exchanging the histidines by alanines improves the binding properties to Si, while the
adsorption strength to GaAs is noticeably reduced. For the randomly permuted sequence, the binding strength 
to GaAs is left widely unaffected, while binding to Si is as strong as to GaAs. 

Employing multicanonical (MUCA) Monte Carlo sampling~\cite{BergNeu,Janke98}, we analyze single-molecule folds 
in the bulk and illuminate the thermodynamic
and structural properties of these peptides. Generalised-ensemble methods applied to all-atom 
descriptions of proteins have been very successful in the past, e.g., in revealing 
the statistical mechanics in the folding process of small proteins~\cite{Gg1,Gg2,HaTc}. 
For sequences with more than 20 residues, studies of thermodynamics and kinetics 
employing realistic physical models are computationally extremely demanding. For such systems, 
reduced all-atom models~\cite{irb1} or models at a higher coarse-grained level~\cite{dill1,still1} 
could be, depending on the particular question, much more promising.
Coarse-grained lattice and off-lattice models allow for systematic thermodynamic studies and,
at least partly, sequence analyses of heteropolymers with up to 100 monomers~\cite{bj1,bj2,MbHaWj}. 

Firstly, we have simulated all three molecules in vacuum. Then, in order to see the effects of solvation, we
have also performed extensive simulations of the commonly used surface-accessible area solvent 
model with OONS atomic solvation parameter
set~\cite{Ooi}. The preferential properties of this parameter set compared to others were 
reported in previous works~\cite{Masuya,Berg}. 

The rest of the paper is organized as follows. After the description of the peptide model and the computational 
methods used in this study in Sect.~\ref{secmet}, 
we discuss in Sect.~\ref{secsol} exemplified effects of solvation compared with results obtained in the vacuum
simulations. In Sect.~\ref{sechc}, we discuss in detail the helix-coil transitions for the three peptides in solvent
by means of fluctuations of several energetic and structural quantities. Section~\ref{secfree} addresses the
folding channels in the free-energy landscape. The paper concludes with a summary in
Sect.~\ref{secsum}. 
%
\section{Peptide Model and Simulation Method}
\label{secmet}
\subsection{The Peptide Model}
In our simulations a peptide is modeled with all of its atoms. Each atom $i$, located
at the position ${\bf r}_i$, carries a partial charge $q_i$. Covalent bonds between atoms, according to the
chemical structure of the amino acids, are considered rigid, i.e., bond lengths are kept constant, as well as
bond angles between covalent bonds and certain rigid torsion angles. Distances between nonbonded atoms $i$ and $j$ 
are defined as $r_{ij}=|{\bf r}_i-{\bf r}_j|$ and measured in \AA\/ in the following. The set of degrees of freedom 
covers all dihedral torsion angles $\boldsymbol{\xi}=\{\xi_\alpha\}$ of $\alpha$th residue's backbone 
($\phi_\alpha$, $\psi_\alpha$, $\omega_\alpha$) and side chain ($\boldsymbol{\chi}=\chi_\alpha^{(1)},\chi_\alpha^{(2)},\ldots$).
The model incorporates electrostatic Coulomb interactions between the partial atomic charges (all energies in kcal/mol),
\begin{equation}
\label{eqc}
E_{\rm C}(\boldsymbol{\xi})  =  332 \sum_{i,j} \frac{q_i q_j}{\varepsilon r_{ij}(\boldsymbol{\xi})},
\end{equation}
effective atomic dipole-dipole interaction modeled via Lennard-Jones potentials~\cite{rem1},
\begin{equation}
\label{eqlj}
E_{\rm LJ}(\boldsymbol{\xi}) =  \sum_{i,j} \left( \frac{A_{ij}}{r^{12}_{ij}(\boldsymbol{\xi})} - 
\frac{B_{ij}}{r^6_{ij}(\boldsymbol{\xi})} \right),
\end{equation}
O-H and N-H hydrogen-bond formation, 
\begin{equation}
\label{eqhb}
E_{\rm HB}(\boldsymbol{\xi})  =  \sum_{i,j} \left( \frac{C_{ij}}{r^{12}_{ij}(\boldsymbol{\xi})} - \frac{D_{ij}}{r^{10}_{ij}(\boldsymbol{\xi})} \right),
\end{equation}
and considers dihedral torsional barriers (if any):
\begin{equation}
\label{eqtor}
E_{\rm tor}(\boldsymbol{\xi}) =  \sum_l U_l \left( 1 \pm \cos (n_l \xi_l ) \right). 
\end{equation}
The total energy of a conformation, whose structure is completely defined by the set of dihedral angles
$\boldsymbol{\xi}$, is 
\begin{equation}
\label{ECEPP3}
E_0(\boldsymbol{\xi}) =  E_{\rm C}(\boldsymbol{\xi}) + E_{\rm LJ}(\boldsymbol{\xi}) + E_{\rm HB}(\boldsymbol{\xi}) + 
E_{\rm tor}(\boldsymbol{\xi}).
\end{equation}
The parameters $q_i, A_{ij},B_{ij}, C_{ij}, D_{ij}, U_l$, and $n_l$ 
are taken from the ECEPP/3 ({\bf E}mpirical {\bf C}onformational {\bf E}nergies 
for {\bf P}roteins and {\bf P}olypeptides) force field~\cite{ecepp}, one of the most commonly used all-atom force fields.
In all simulations the dielectric constant was set to $\varepsilon=2$, which is the vacuum value. Proline's
$\phi$ is considered rigid at $-68.8^\circ$. We always used the trans down-puckering conformation of the proline ring.  
For the implicit-solvent simulations, the model is extended by the solvation-energy contribution, which 
is given by~\cite{Solvterm}
\begin{equation}
\label{eqsol}
E_{\rm solv}(\boldsymbol{\xi})=\sum\limits_i\sigma_iA_i(\boldsymbol{\xi}),
\end{equation}
where  $A_i$ is the solvent-accessible surface area of the $i$th atom for a given conformation and $\sigma_i$
is the solvation parameter for the $i$th atom. The values for $\sigma_i$ depend on the type of the $i$th atom and 
are parameterized according to the suggestions given in Ref.~\cite{Ooi}. The total potential energy of the molecule then reads 
\begin{equation}
\label{eqtot}
E_{\rm tot}(\boldsymbol{\xi}) = E_0(\boldsymbol{\xi}) + E_{\rm solv}(\boldsymbol{\xi}).
\end{equation}
The described peptide model and the ECEPP/3 parameterization is implemented
in the software package SMMP~\cite{SMMP}, which we used for our study.

In Table~\ref{gemtable}, we have listed the three sequences investigated with this model. 
\begin{table}
\caption{\label{gemtable} Peptide sequences studied in this work.}
\begin{tabular}{c|c}\hline\hline
S1 & AQNPSDNNTHTH\\ 
S2 & AQNPSDNNTATA\\
S3 & TNHDHSNAPTNQ\\ \hline\hline
\end{tabular}
\end{table}
\subsection{Multicanonical Sampling}
Multicanonical sampling~\cite{BergNeu,Janke98,berg2} is a generalized-ensemble method, in which
conformations are ideally sampled according to a flat energy distribution $p_{\rm muca}(E)={\rm const}$.,
i.e., the Markovian dynamics of the algorithm corresponds to a random walk in energy space.
The desired canonical distribution at a certain temperature $T$ is given by
$p_{\rm can}(E,T)\sim n(E)\exp\left(-E/RT \right)$, where $n(E)$ is the density
of states and the gas constant takes the value $R\approx 1.99\times 10^{-3}$ kcal/K\,mol in
the units used in this paper.
Since canonical and multicanonical energy distributions are trivially related via 
$p_{\rm can}(E,T)\sim W_{\rm muca}^{-1}(E)\exp\left(-E/RT \right)p_{\rm muca}(E)$, the main task is a precise
determination of the multicanonical weights $W_{\rm muca}(E)\sim n^{-1}(E)$.

The implementation of MUCA is not straightforward as the density of states  $n(E)$ is unknown {\it a priori}.
Therefore, the weights $W_{\rm muca}(E)$ have to be determined in the
first stage of the simulation process by an iterative procedure until
the multicanonical histogram $H(E)\approx {\rm const.}$ in the desired energy interval. 
An efficient, error-weighted estimation method for the multicanonical weights is
described in detail in Refs.~\cite{Janke98,berg2}. We note that the efficiency of the determination
of the multicanonical weights usually depends on the choice of the simulation temperature, which was
in the present study $T_{\rm sim}=1\,000 K$. The reason is that, since the ``flat'' energy histogram covers a larger region
in subsequent recursions, energetic states are hit for the first time, where the multicanonical
weights are still undetermined. Because the ratio of the weights controls the acceptance of a conformational update,
the dynamics of the recursion part of the algorithm is noticeably influenced. This behavior can be ``smoothed'' by a careful choice 
of the simulation temperature. 

Eventually, in the second stage of the multicanonical simulation, a long production run is performed based on fixed 
multicanonical weights. Since the weights for all energetic states in the desired energy range have already been determined,
the choice of a certain simulation temperature is unnecessary.

In our concrete implementation, we first carried out Metropolis simulations 
at relatively high simulation temperatures and MUCA test runs which enabled us 
to determine the required energy range. This interval was then divided into
bins of $1\,$kcal/mol.  At each update step, a trial conformation was obtained by changing 
a dihedral angle $\xi_i\to \xi'_i$ within the range $[-180^\circ,180^\circ]$, which
was accepted according to the transition probability 
$\omega(\boldsymbol{\xi}\to \boldsymbol{\xi'})=
{\rm min}\left[\exp\left(S(E(\boldsymbol{\xi}))-S(E(\boldsymbol{\xi'}))\right),1\right]$,
where $S(E(\boldsymbol{\xi}))=-\log W_{\rm muca}(E(\boldsymbol{\xi}))$
can be identified with the microcanonical entropy.
The dihedral angles were always visited in a predefined, sequential
order, i.e., a sweep is a cycle of $N$ Monte Carlo steps ($N$ = total number of dihedral angles).

\begin{figure}
\centerline{\epsfxsize=8.8cm \epsfbox{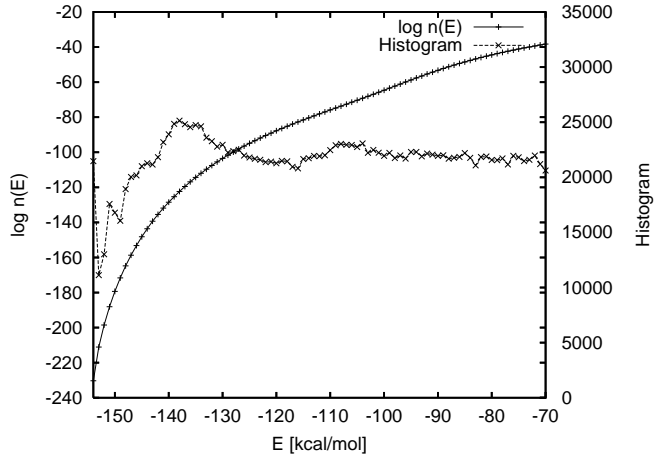}}
\caption{\label{weight} Natural logarithm of the density of states $n(E)$ and multicanonical histogram 
from the simulation of the sequence AQNPSDNNTHTH (S1) in solvent.}
\end{figure}

The weights were built in $200$ recursions during a
long {\it single} simulation, where the multicanonical parameters
were iterated every $10\,000$ sweeps.
Then, we performed a full simulation of two million sweeps with fixed weights, which covers the
temperature region up to $T_{\max}=1\,000\,$K reliably. In Fig.~\ref{weight}, the 
density of states and the multicanonical histogram for the first sequence considered, AQNPSDNNTHTH (S1), is 
shown. As seen from this figure, the multicanonical histogram is indeed ``flat'' which is a necessary condition 
for the multicanonical technique to be reliably working.
Statistical expectation values for any thermodynamic quantity $\cal{A}$ and all temperatures can finally be calculated 
from the time series recorded during the multicanonical production run:
\begin{equation}
\langle{\cal{A}}\rangle =\frac{\sum_t {\cal{A}}(\boldsymbol{\xi}^{(t)}) 
W_{\rm muca}^{-1}(E(\boldsymbol{\xi}^{(t)})) e^{-\beta E(\boldsymbol{\xi}^{(t)})}}
{\sum_t W_{\rm muca}^{-1}(E(\boldsymbol{\xi}^{(t)})) e^{-\beta E(\boldsymbol{\xi}^{(t)})}},
\end{equation}
where $\boldsymbol{\xi}^{(t)}$ labels the conformation at ``time'' $t$ and $\beta=1/RT$ is the inverse thermal
energy. 

The derivative of the quantity  $\langle\cal{A}\rangle$ with respect to the thermal energy is given by
\begin{equation}
\frac{d \langle {\cal{A}} \rangle}{d(RT)} = 
\frac{1}{(RT)^2}\left(\langle E{\cal{A}}\rangle - \langle E \rangle \langle {\cal{A}} \rangle\right) .
\end{equation}
Expressions like this are typically used to discuss the influence of thermal fluctuations on $\cal{A}$.
\begin{figure}
\centerline{\epsfxsize=8.8cm \epsfbox{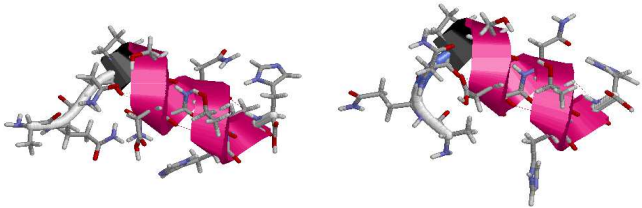}}
\caption{\label{fig:s1ref} (Color online) Low-energy reference conformations of the peptide S1 in
vacuum (left) and solvent (right) for the calculation of the overlap parameter~(\ref{eqov}).}
\end{figure}
\begin{figure}
\centerline{\epsfxsize=8.8cm \epsfbox{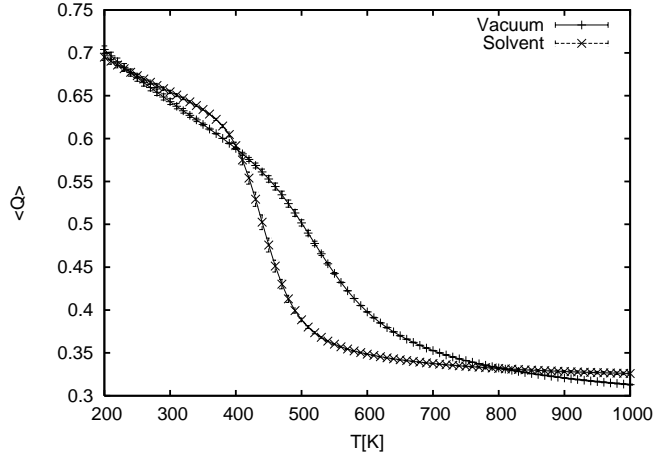}}
\caption{\label{xxx2} Average angular overlap $Q$ between ensemble conformations and the reference conformations
in Fig.~\ref{fig:s1ref} for the peptide S1 as a function of temperature in vacuum and solvent.}
\end{figure}
%
\section{Solvation Effects}  
\label{secsol}
In this section, we compare the folding behavior of the exemplified peptide S1 in vacuum and solvent, respectively. 
Changes of energy fluctuations, i.e., the widths of the energy distributions, signalize typically a crossover or transition
between significantly different macrostates (``phases'') of the system considered. For polymers or peptides, the crossover 
between such macrostates is accompanied by a cooperative conformational transition. Thus it is reasonable to compare
the behavior of the peptide in vacuum and solvent with regard to energetic fluctuations, the specific heat (in units of $R$)
\begin{equation}
C_V =  \frac{1}{(RT)^2} \left(\langle E^2 \rangle - \langle E \rangle^2\right),
\end{equation}
and with respect to the angular overlap parameter~\cite{HaOk99,HaOn,berg3} as a structural quantity (``order'' parameter),
which is suitably defined as
\begin{equation}
\label{eqov}
Q(\boldsymbol{\xi}^{(t)},\boldsymbol{\xi}^{\rm ref})=1-\frac{1}{90^\circ N} \sum^{N}_{i=1}  d(\xi_i^{(t)}, \xi_i^{\rm ref}),
\end{equation}
where $d(\alpha, \alpha')={\rm min}\,(|\alpha-\alpha'|, 360^\circ-|\alpha-\alpha'|)$. In this expression, the dihedral angles
of the actual conformation $\boldsymbol{\xi}^{(t)}$ are compared with the corresponding torsion angles of 
a suitable reference conformation, $\boldsymbol{\xi}^{\rm ref}$.  
\begin{figure}
\centerline{\epsfxsize=8.8cm \epsfbox{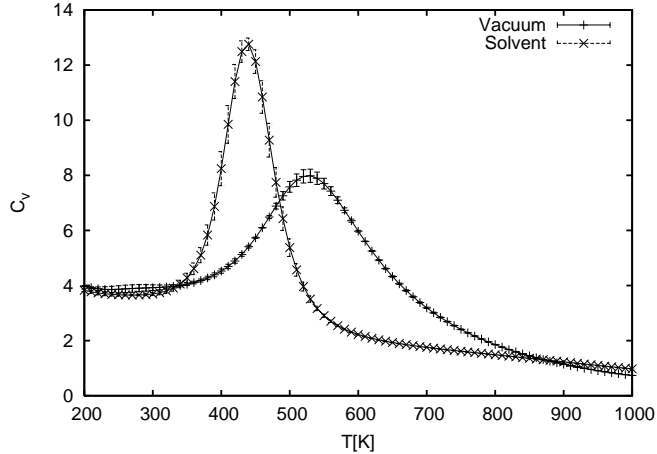}}
\caption{\label{cv2} Specific heat vs.\ temperature for the peptide S1 in vacuum and solvent. }
\end{figure}
\begin{figure}
\centerline{\epsfxsize=8.8cm \epsfbox{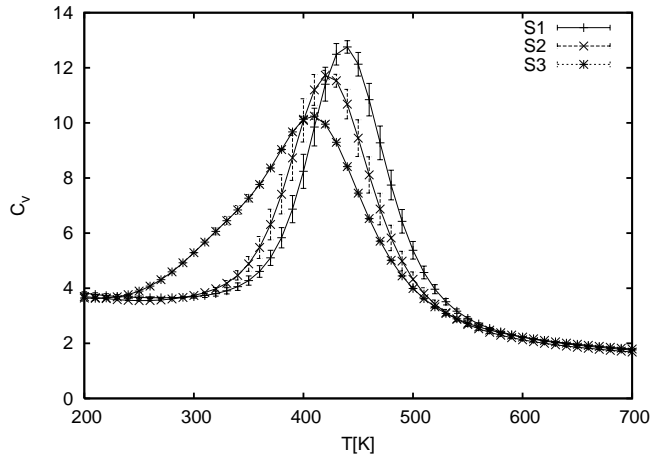}}
\caption{\label{fig:cv} Specific heat vs.\ temperature for the three peptides in solvent.}
\end{figure}
\begin{figure}
\centerline{\epsfxsize=8.8cm \epsfbox{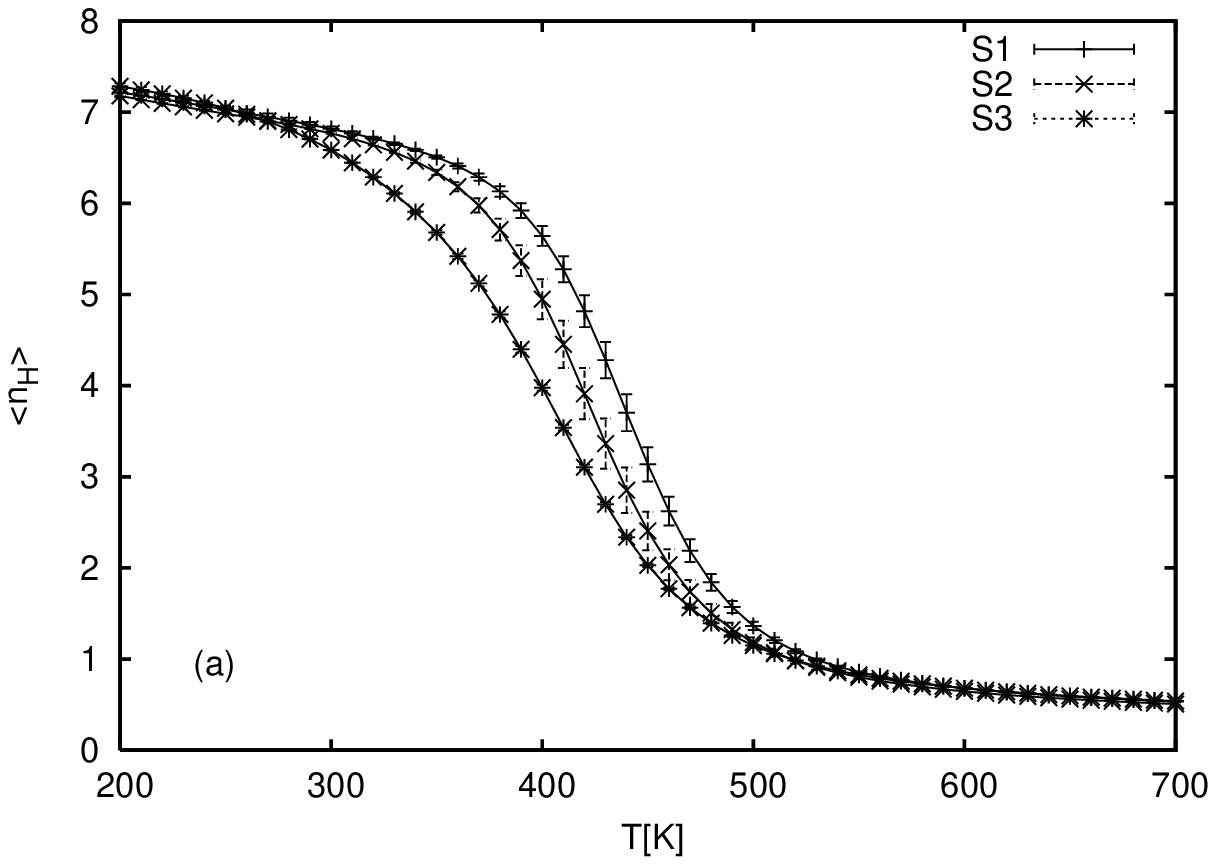}}
\centerline{\epsfxsize=8.8cm \epsfbox{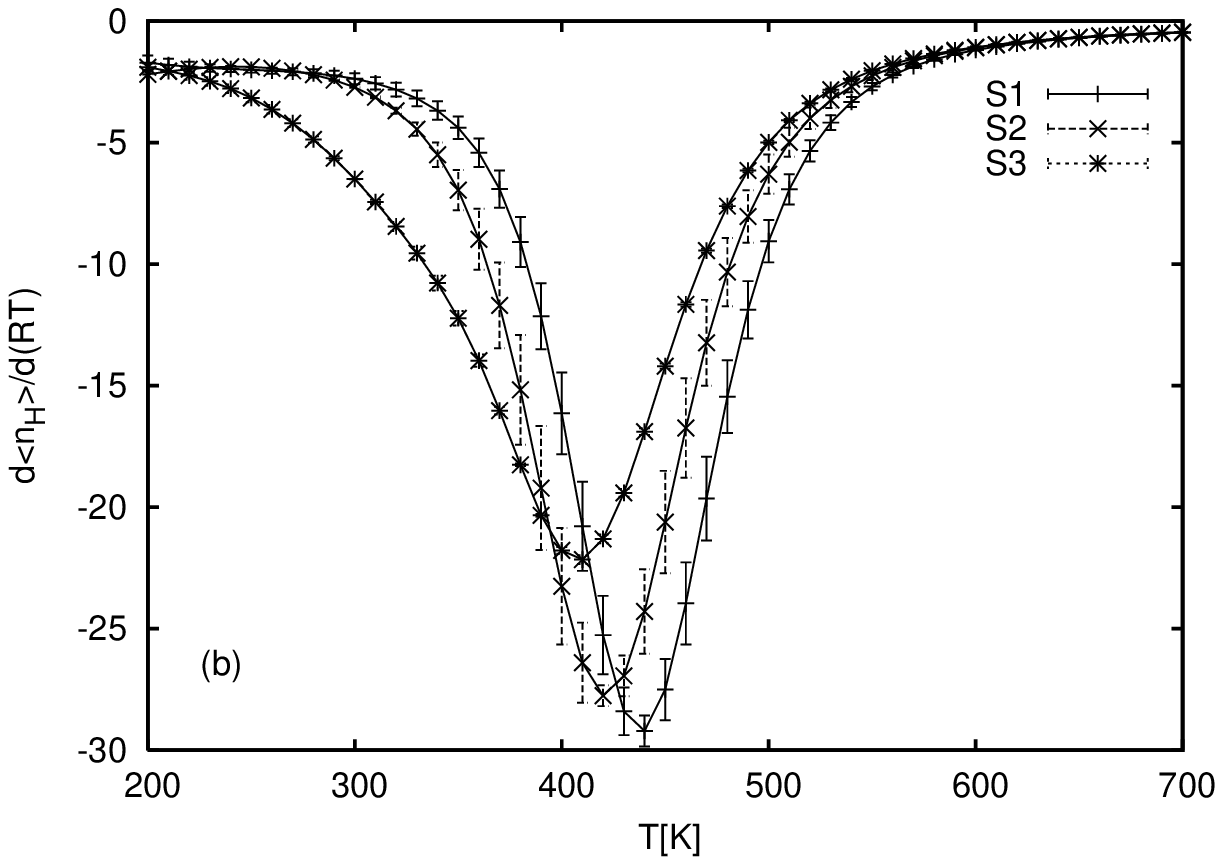}}
\caption{\label{fig:hel} Average number of helical residues (a) and derivative (b) vs.\ temperature 
for the systems in solvent.}
\end{figure}
Since the overlap parameter (\ref{eqov}) is a measure for the similarity of any conformation and the reference 
conformation, it can be considered as a system state parameter: $Q=1$ only if the considered conformation is 
identical with the reference conformation, which is here chosen to be a typical representative
of the (low-energy) helical phase. In Fig.~\ref{fig:s1ref}, the helical low-energy reference conformations of sequence
S1 in vacuum (left) and solvent (right) are shown. The main differences regard the side-chains
and the tail, whereas the helical part is hardly influenced by solvation effects. The overlap $Q$ between
these two conformations comparing {\em all} dihedral angles is 0.595, while considering only the backbone dihedrals, 
$Q\approx 0.690$. In the latter case, 
the comparatively still small coincidence is mainly due to the non-helical tails, which are highly flexible,
i.e., there are no stabilizing hydrogen bonds with the backbone. Proline breaks the helix in both cases.

In Fig.~\ref{xxx2} we have plotted the average overlap parameter $\langle Q\rangle$ for this exemplified sequence
in vacuum and solvent. For low temperatures, in both cases most of the conformations in the ensemble have similarities
with the reference conformations, i.e., the ensemble is dominated by helical conformations. For the peptide in solvent,
the average overlap parameter decreases rapidly at about 440 K, the conformations in the high-temperature phase
are random coils. The situation is comparable for the peptide in vacuum, with the noticeable differences that
$\langle Q\rangle$ decreases much slower at a transition temperature near 530 K. This is confirmed by 
considering the energetic fluctuations of the system, i.e., the specific heat as shown in Fig.~\ref{cv2}.
In the vicinity of the peak temperatures, the peptides exhibit conformational activity. The helix-coil transition
peak is stronger and sharper for the peptide in solvent, the transition temperatures are close to the
above-mentioned values.

Summarizing, the main effect of the solvent is the strengthening of the helix-coil transition which is 
also present in the vacuum case. Furthermore, the transition temperature is shifted by about 100 K
towards lower temperatures. These results are as expected, since it is known that solvent stabilizes
secondary structures and therefore the barrier to resolve the helix is higher than in the vacuum case
and the relaxation of the fluctuations of the peptide-solvent coupling degrees of freedom leads to
a lower transition temperature. These differences have also been observed in studies using other 
parameter sets~\cite{SCH}.

It should be noted that also in the OONS implicit-solvent model the transition temperature is probably
still strongly overestimated as is already known from studies of other helical peptides~\cite{hansmann1}.
One of the reasons is the choice of a temperature-independent solvent-peptide coupling strength
and the ``smeared'', nonlocal and static polar environment without intrinsic fluid properties. 
%
\section{Helix-Coil Transitions of the Peptides in Solvent}
\label{sechc}
In the following we discuss the thermodynamic properties of the three synthetic sequences   
employing the ECEPP/3 force field with OONS implicit-solvent parameter set. Considering several 
quantities we find strong indications for helix-coil transitions in all three cases. 
Helix-coil transitions in peptides and nucleic acids were first addressed by 
Zimm and Bragg ~\cite{ZimmBr,Zimm} and have been studied extensively~\cite{PolSch}.
\begin{figure}
\centerline{\epsfxsize=8.8cm \epsfbox{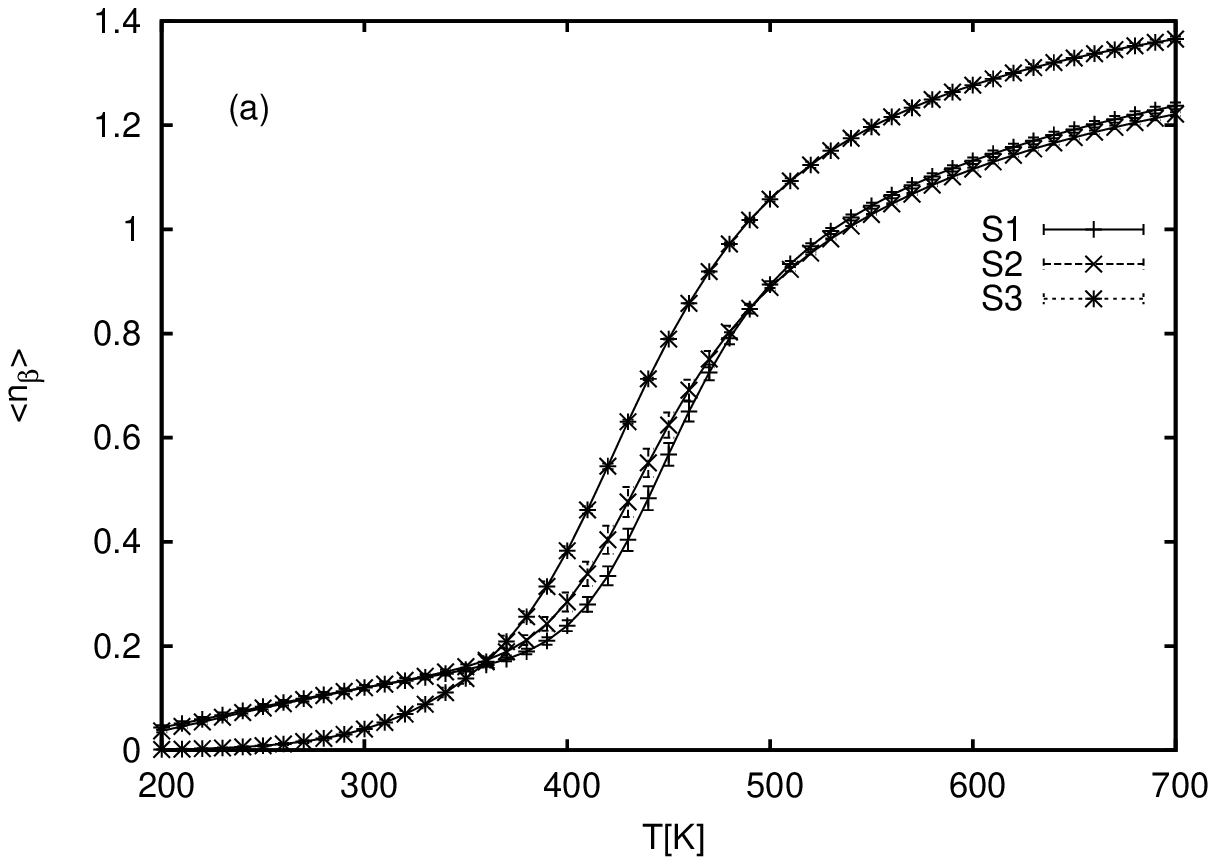}}
\centerline{\epsfxsize=8.8cm \epsfbox{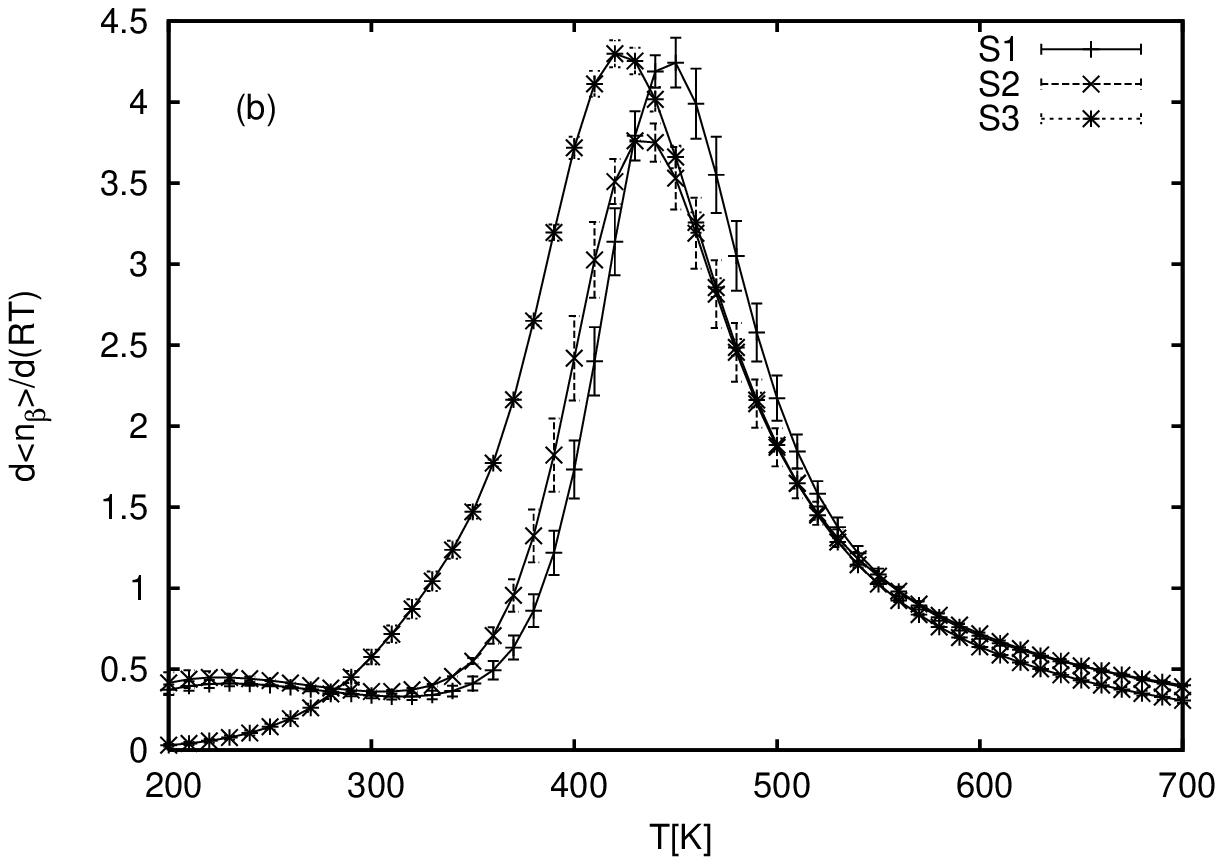}}
\caption{\label{fig:beta} Average number of $\beta$-sheet structures (a) and derivative (b) vs.\ temperature 
for the peptides in solvent.}
\end{figure}

As a first indication for the conformational transitions, we find strong peaks in the specific heat as
shown for the three peptides in solvent in Fig.~\ref{fig:cv}. The peaks are located near 440, 420, and 410 K 
for the wild-type S1, mutant S2, and randomly permuted sequence S3, respectively. 
The character of the conformational transition is identified by measuring the temperature-dependence
of $\alpha$-helicity and $\beta$-sheetness.
The helicity is a natural order parameter for the identification of helix-coil transitions in peptides. A residue is
defined to be in $\alpha$-helical state, if its backbone dihedral angles $\phi$ and $\psi$ are in the range
$(-70 \pm 20^{\circ})$ and $(-37 \pm 20^{\circ})$, respectively.
We have shown the changes of this quantity and its derivative versus temperature in Figs.~\ref{fig:hel}(a) and (b), 
respectively. 

In the calculation of the helicity, end residues are not taken into account, because
end residues are very flexible and do not conform to any definite state.
There is also a proline residue in all chains which is known as helix-breaker, because it lacks a
primary amine group and due to the peptide bond to the preceding amino acid, there is no 
H-atom allowing for the formation of a hydrogen bond that could stabilize an $\alpha$-helix
or a $\beta$-sheet structure. Furthermore, the rigid proline side chain 
typically forces for steric reasons the $\psi$ angle of the preceding amino acid to take non-helical values.
Hence, the maximum number of residues in a helical
segment is eight for the three sequences under consideration. Rapid decreases of the average helicities
(as shown in Fig.~\ref{fig:hel}(a)) are noticed for all three peptides. The transition temperatures
lie in the same temperature region as the peaks of the specific heat, as can be read off from
the fluctuations around the average helicity as plotted in Fig.~\ref{fig:hel}(b).

In the force-field parameterization used, $\beta$ sheets can be excluded in the low-temperature region. 
This is confirmed by the
plots in Figs.~\ref{fig:beta}(a) and (b), where the sheetness $\langle n_\beta\rangle$, 
which is the average number of Ramachandran angles in a $\beta$-sheet state 
[i.e., $\phi\in (-150 \pm 30^{\circ})$ and $\psi\in(150 \pm 30^{\circ})$], and the fluctuations of this quantity 
are shown. No noticeable 
$\beta$-sheet structure is identified in the low-temperature region because the whole ensemble consists of strongly 
helical conformations. The average sheetness increases slightly above the conformational-transition   
temperature, but this signal is relatively weak and the high-temperature ensemble is expected to be
dominated by random-coil structures.

\begin{figure}
\centerline{\epsfxsize=8.8cm \epsfbox{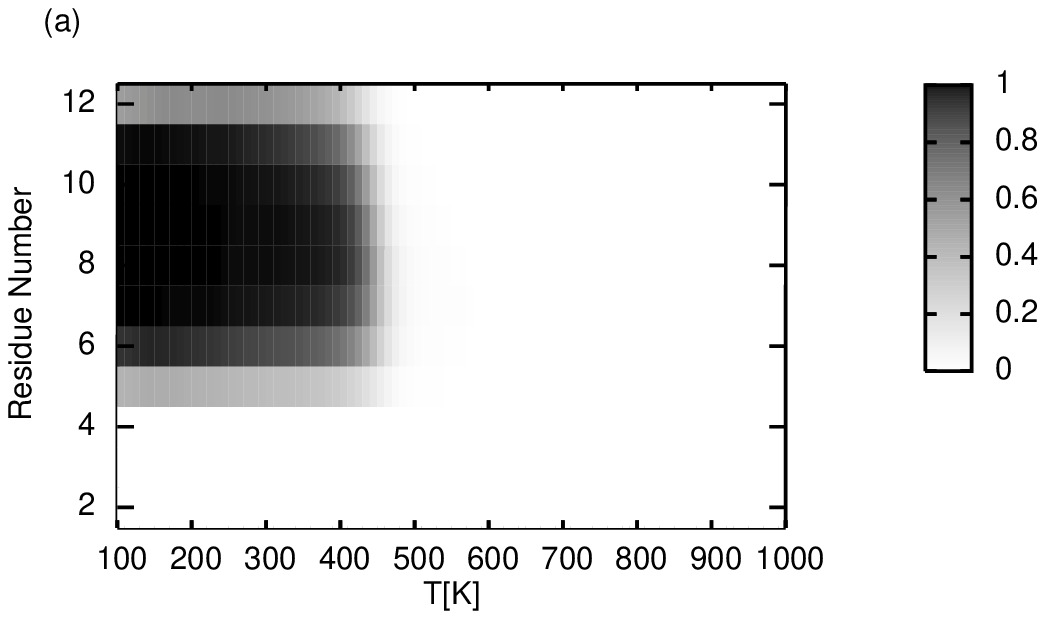}}

\vspace*{-3mm}
\centerline{\epsfxsize=8.8cm \epsfbox{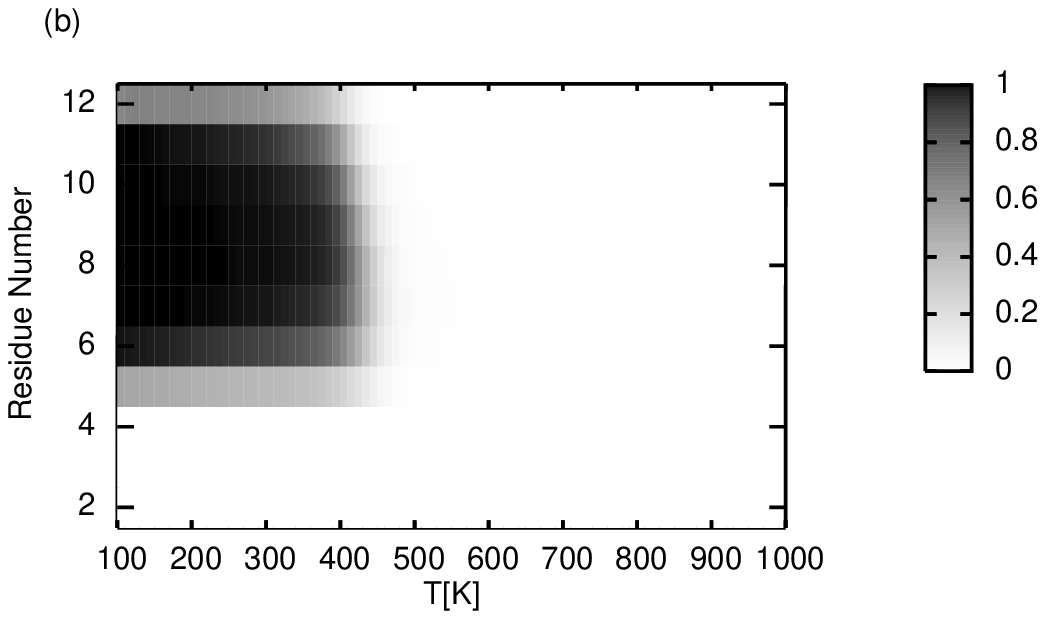}}

\vspace*{-3mm}
\centerline{\epsfxsize=8.8cm \epsfbox{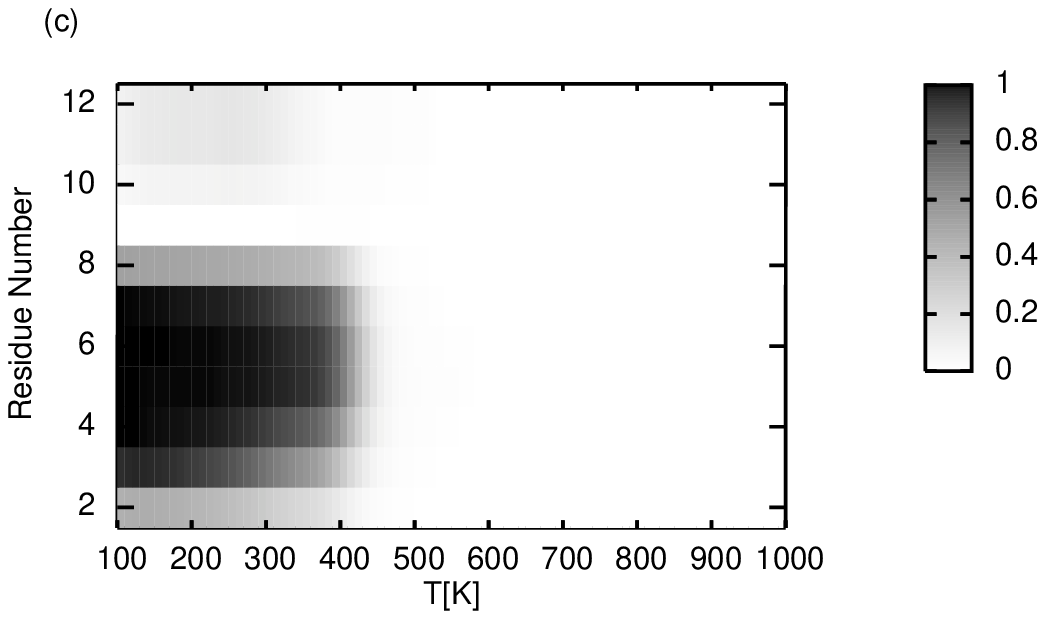}}
\caption{\label{fig:segment} Probability profiles of each residue to be a part of a 3-residue helical 
segment as a function of temperature for (a): S1, (b): S2, and (c): S3 in solvent.}
\end{figure}

Another interesting quantity is the probability of the individual residues to become ``helical''. 
In Figs.~\ref{fig:segment}(a)--(c), we have plotted the color-coded profiles of the probabilities
of each residue to be in a helical state. Since it is not intuitive that a single residue 
can form an $\alpha$-helix motif, although its dihedral angles are in the range of 
$\alpha$-helical region of the Ramachandran map, we define a residue to be helical only if
it is part of a helical segment with at least three successive helical pairs of Ramachandran angles. 
This allows a clearer view on the helix formation. For the three considered peptides in implicit 
solvent, the helix-coil transition is a sharp one-step process in the temperature region between 400 
and 500 K. The single helical segment of the peptides S1 and S2 is formed by the residues 5 to 12
(counting from the N terminus). Proline at position 4 in the sequence breaks the helix and 
the 1-4 residual tail is coil-like. Surprisingly, the randomly permuted sequence S3, where proline is located
at position 9 (which is, unfortunately, also at the fourth position -- counted from the C terminus), exhibits
in addition to the strong 1-8 helix a second helical segment between residues 10 and 12. Although
this signal is weak, a non-negligible subset of conformations in the low-temperature ensemble
contains {\em two} independent helical segments, broken by proline. The transition temperature 
for the 10-12 helix is slightly smaller than for the main segment and lies below 400 K. Note that
the formation of the second helix for sequence S3 can also be observed as ``shoulder'' in the 
corresponding specific heat in Fig.~\ref{fig:cv}.  
\begin{figure}
\centerline{\epsfxsize=8.8cm \epsfbox{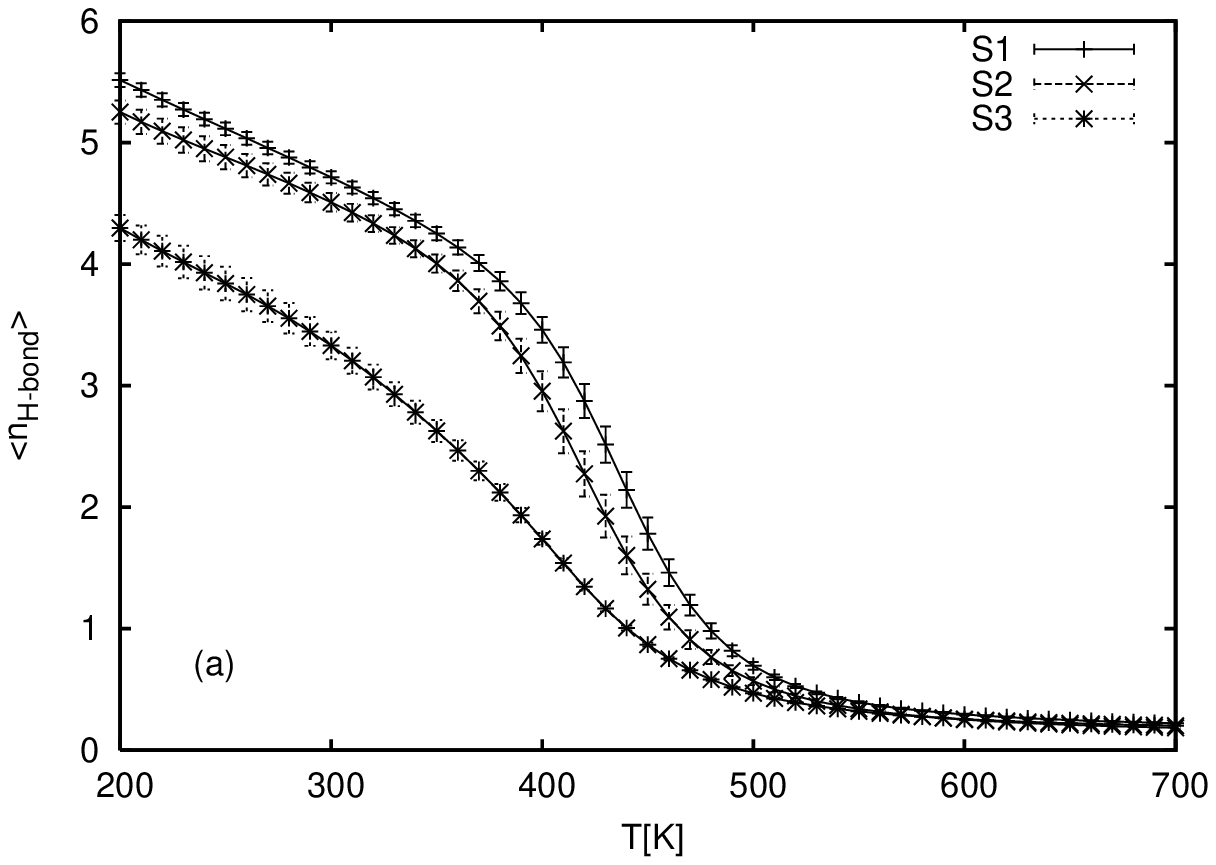}}
\centerline{\epsfxsize=8.8cm \epsfbox{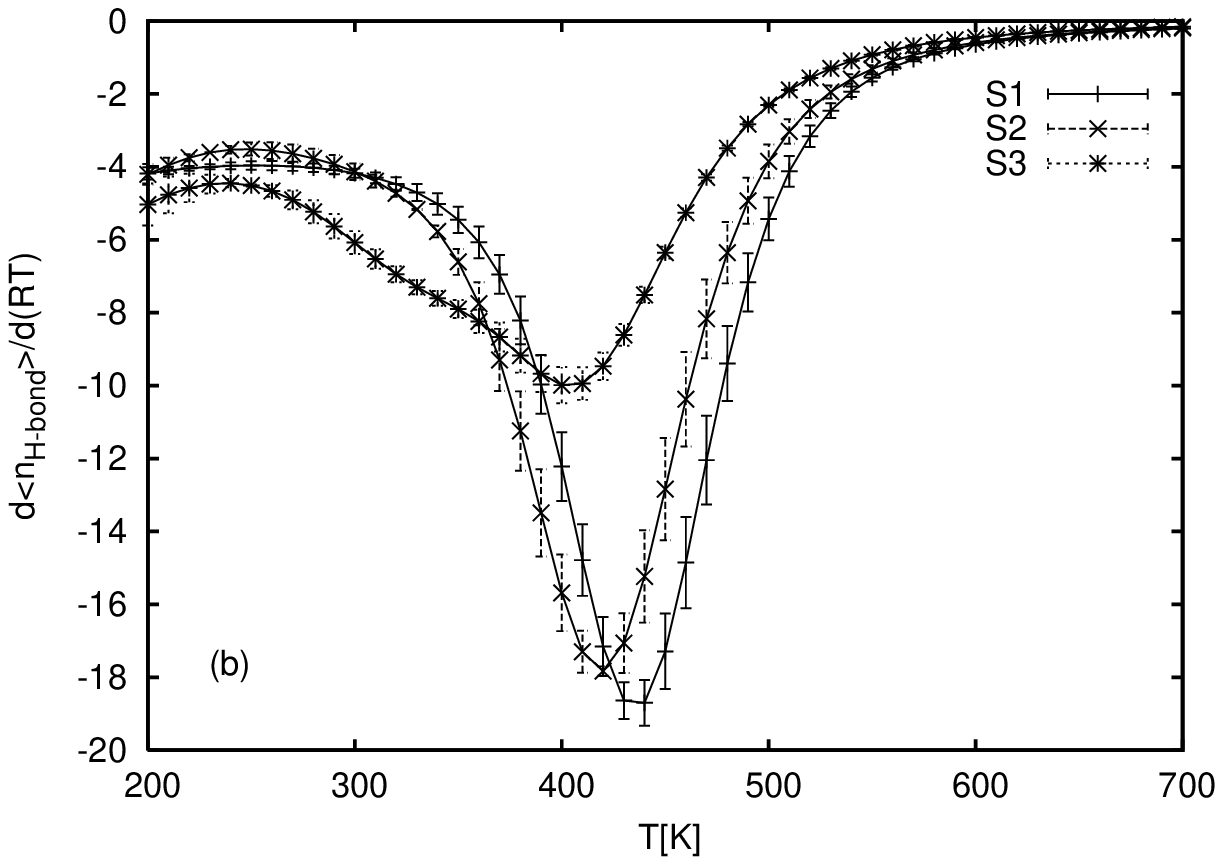}}
\caption{\label{fig:hbond} Average number of hydrogen bonds (a) and derivative (b) vs.\ temperature 
for the three peptides in solvent.}
\end{figure}

Hydrogen bonds are mainly responsible for the formation and stability of secondary structures such as $\alpha$-helices.
In Figs.~\ref{fig:hbond}(a) and (b), the average numbers of hydrogen bonds $\langle n_{\rm H-bond}\rangle$ and 
their fluctuations, respectively, are shown for the three sequences. 
In the helical phase, conformations typically possess approximately 
4 -- 6 hydrogen bonds on average. Hydrogen bonds of the 
chains with water are obviously not counted employing an implicit peptide-solvent model and averages shown in 
the figure reflect only intrinsic hydrogen bonds. As can be seen in Fig.~\ref{fig:hbond}(a), 
the peptide with the randomly
permuted sequence S3 behaves noticeably different than sequences S1 and S2. For S3, $\langle n_{\rm H-bond}\rangle$
decreases more smoothly with increasing temperature, i.e., the breaking of the individual hydrogen bonds is a process
of relatively weak cooperativity. In this case, the hydrogen bonds are comparatively weak and the two 
independent helical segments and, more globally, the helical phases, are not very stable, as the
fluctuations show in Fig.~\ref{fig:hbond}(b). This effect is mainly due to the position of the proline in the
chain. Although its distance from the ends is identical in all three sequences, the asymmetry with 
respect to the dihedral constraints destabilizes the larger helical segment of S3 (whose average length 
is smaller than the 
helices of S1 and S2) and leads to a noticeable probability of forming a second small and weak helical segment. 

Finally, we have also calculated the radius of gyration as a rather global geometrical quantity,
which is mainly useful for quantifying the structural collapse caused by a conformational
transition. In contrast to the $\Theta$ collapse transition of polymers between non-structured globular and
random-coil conformations, a helix-coil transition is rather a crossover from non-structured conformations to 
conformations with highly ordered segments (helices). For this reason, the gyration radius is a too rough
measure for the order in the helical phase and is therefore of less
importance for the understanding of secondary-structure formation and cannot be deduced from the short-range 
interactions~\cite{Doniach,Zagrovic}.  
\begin{figure}
\centerline{\epsfxsize=8.8cm \epsfbox{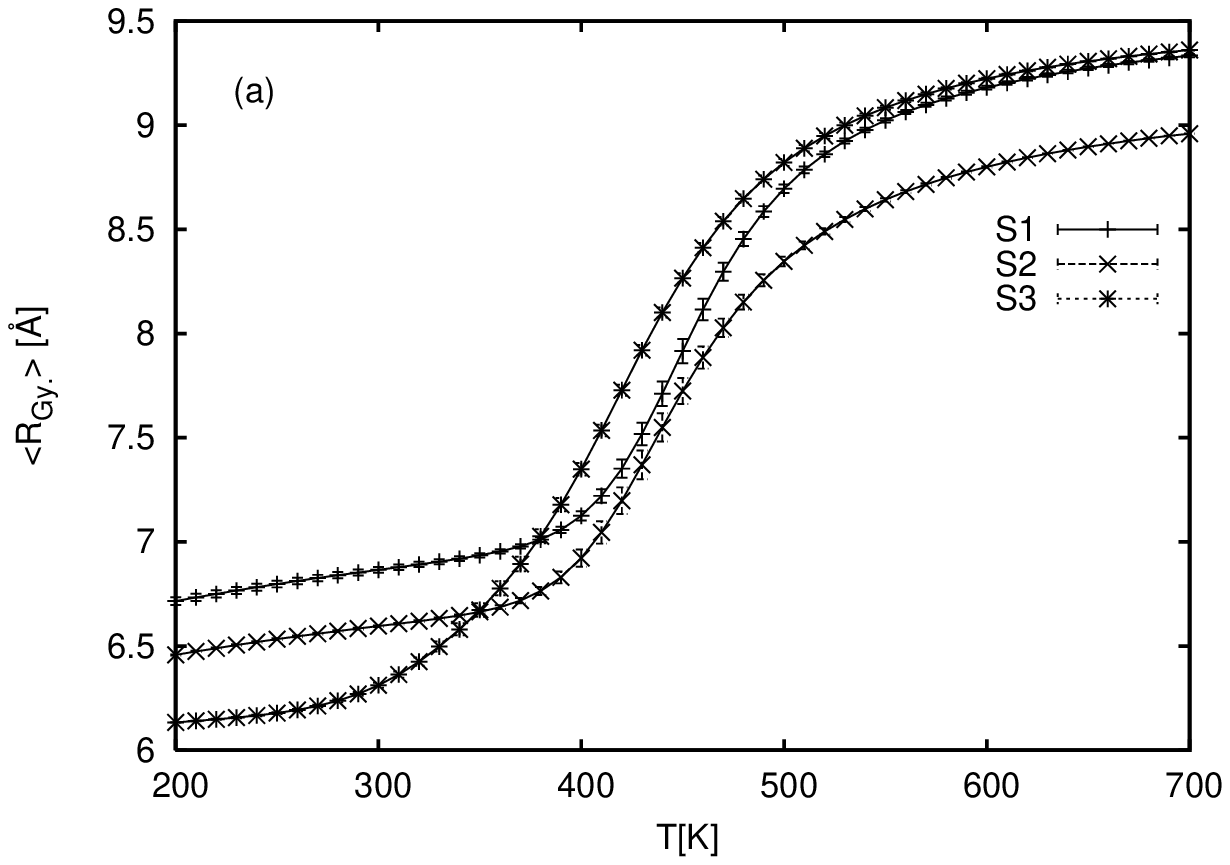}}
\centerline{\epsfxsize=8.8cm \epsfbox{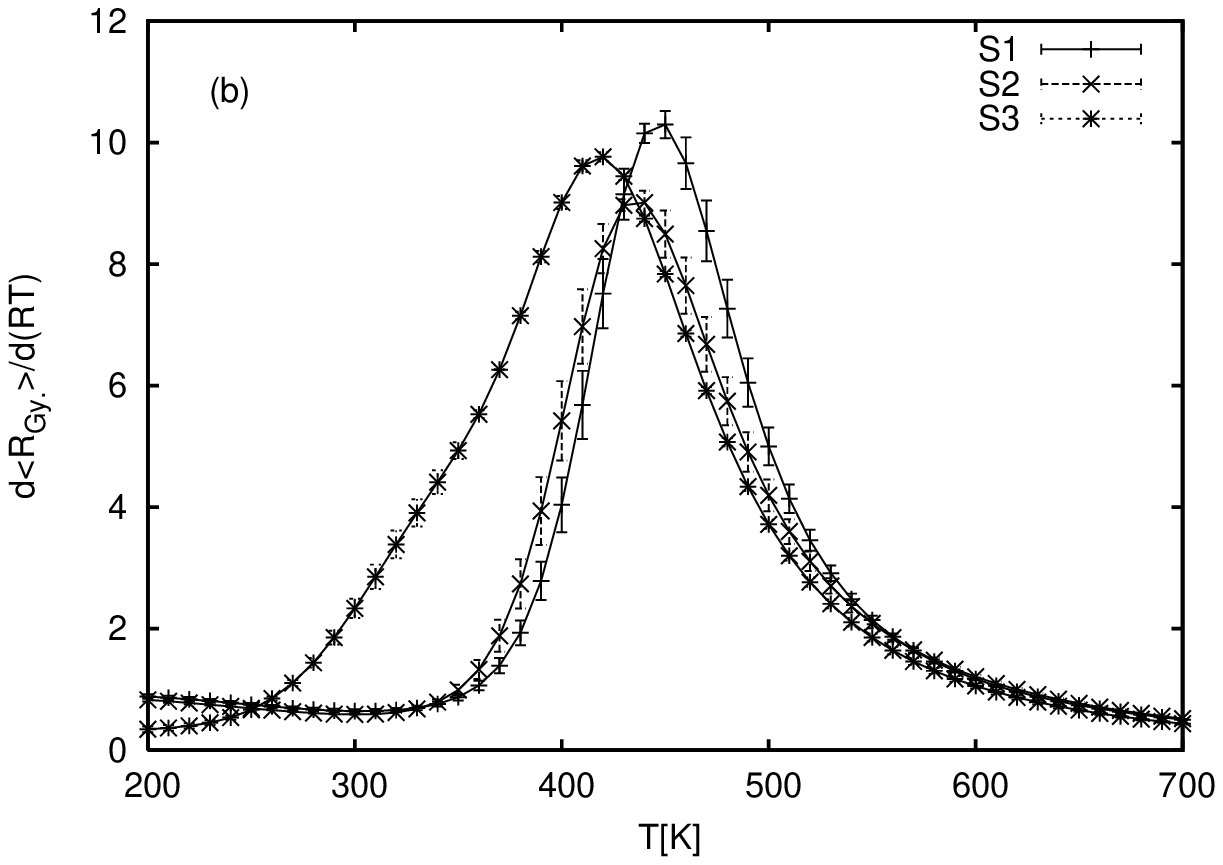}}
\caption{\label{fig:rgy} Average radius of gyration (a) and derivative (b) vs.\ temperature 
for the systems in solvent.}
\end{figure}

In Figs.~\ref{fig:rgy}(a) and (b), the average radius of gyration and its fluctuations 
are shown. Being a measure of the compactness of the molecule, small values of the gyration radius indicate 
more tight-packed structures. For all sequences, the average radius of gyration changes from 
6.0-6.5 {\AA} to 9.0-9.5 {\AA}. The peaks in the 
derivative of  $R_{\rm Gy}$ indicate slightly higher conformational transition temperatures than the 
identified temperatures from fluctuations of energy and helicity. Sequence S3 possesses 
the most compact conformations in the helical phase -- although the average number of hydrogen bonds is 
the smallest, as can also been seen in Table~\ref{tab:delta}, where the differences between maximal and minimal values 
of average radius of gyration and mean number of hydrogen bonds are given. 
\begin{table}
\caption{\label{tab:delta} Differences between maximal and minimal values of 
average gyration radius 
$\Delta \langle R_{\rm Gy}\rangle = \langle R_{\rm Gy}\rangle_{\rm max}-\langle R_{\rm Gy}\rangle_{\rm min}$ 
and mean number of hydrogen bonds 
$\Delta \langle n_{\rm H-bond}\rangle = \langle n_{\rm H-bond}\rangle_{\rm max}-\langle n_{\rm H-bond}\rangle_{\rm min}$ 
for each sequence over the whole temperature range.} 
\begin{tabular}{c|cc}\hline\hline
& $\Delta \langle R_{\rm Gy}\rangle$ [\AA] & $\Delta \langle n_{\rm H-bond}\rangle$ \\ \hline
S1 & 2.6 & 5.4 \\
S2 & 2.5 & 5.1 \\
S3 & 3.2 & 4.1\\
\hline\hline
\end{tabular}
\end{table}

In Table~\ref{tab:hel_coil}, we have listed the peak temperatures identified from fluctuations of
several thermodynamic quantities discussed in this section for the three peptides S1, S2, and S3.
The peak temperatures of the peptides in solvent are compared with the corresponding transition
temperatures identified for the systems in vacuum. While for the peptides in solvent the 
transition temperatures are relatively independent of the fluctuations considered, the deviations
for the vacuum systems are noticeable. This is not surprising, as the systems in solvent are
stabilized by the environment and the helix-coil transition is a cooperative effect that is 
accompanied by a strong coupling to the solvent. In vacuum, the conformational freedom is much larger,
and the helix-coil transition rather an entropic effect. In this case, the finiteness of the systems
is more influential than for the peptides in solvent. 
\begin{table}
\caption{\label{tab:hel_coil} Helix-coil transition temperatures $T_C$ in K, read off for the three sequences 
S1, S2, and S3 in vacuum and solvent from the various quantities 
discussed in the paper. The errors are in the range $\pm 10$ $K$.}
\begin{tabular}{ll|ccccc}\hline\hline
& & $C_V$ & $d\langle R_{\rm Gy}\rangle/dT$ & $d\langle n_{\rm H}\rangle/dT$ & 
$d\langle n_{\rm H-bond}\rangle/dT$ & $d\langle n_{\beta}\rangle/dT$ \\
\hline
 S1 & Vac. & 530 & 600 & 520 & 510 & 550 \\
 & Solv. & 440 & 450 & 440 & 440 & 450 \\
\hline
S2 & Vac. & 520 & 580 & 510 & 510 & 550 \\
 & Solv. & 420 & 440 & 420 & 420 & 430 \\
\hline
S3 & Vac. & 480 & 580 & 460 & 460 & 520 \\
 & Solv. & 410 & 420 & 410 & 400 & 420 \\
\hline\hline
\end{tabular}
\end{table}
\begin{figure}
\centerline{\epsfxsize=8.8cm \epsfbox{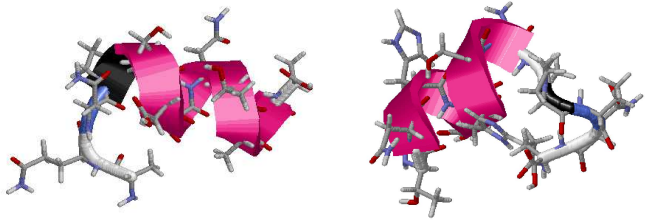}}
\caption{\label{fig:s23ref} (Color online) Lowest-energy reference conformations of sequence S2 (left) 
and S3 (right) in solvent.}
\end{figure}
%
\section{Multicanonical Histograms and Free-Energy Landscapes}
\label{secfree}
\begin{figure}
\centerline{\epsfxsize=8.6cm \epsfbox{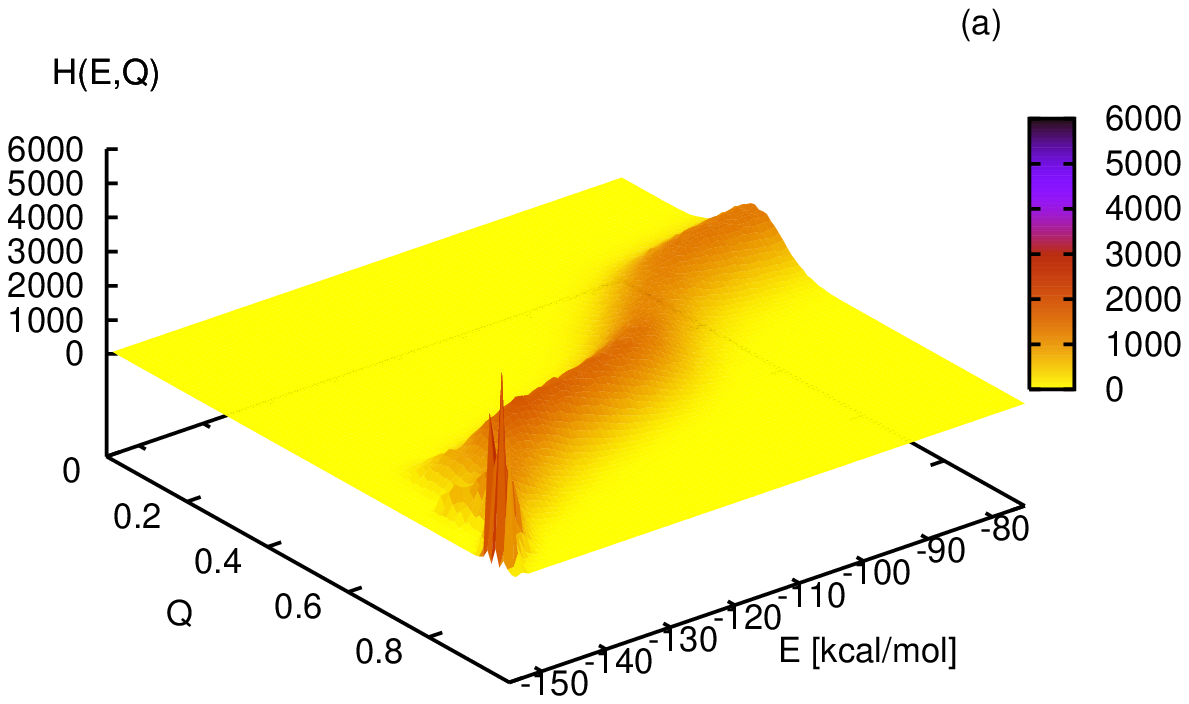}}
\centerline{\epsfxsize=8.6cm \epsfbox{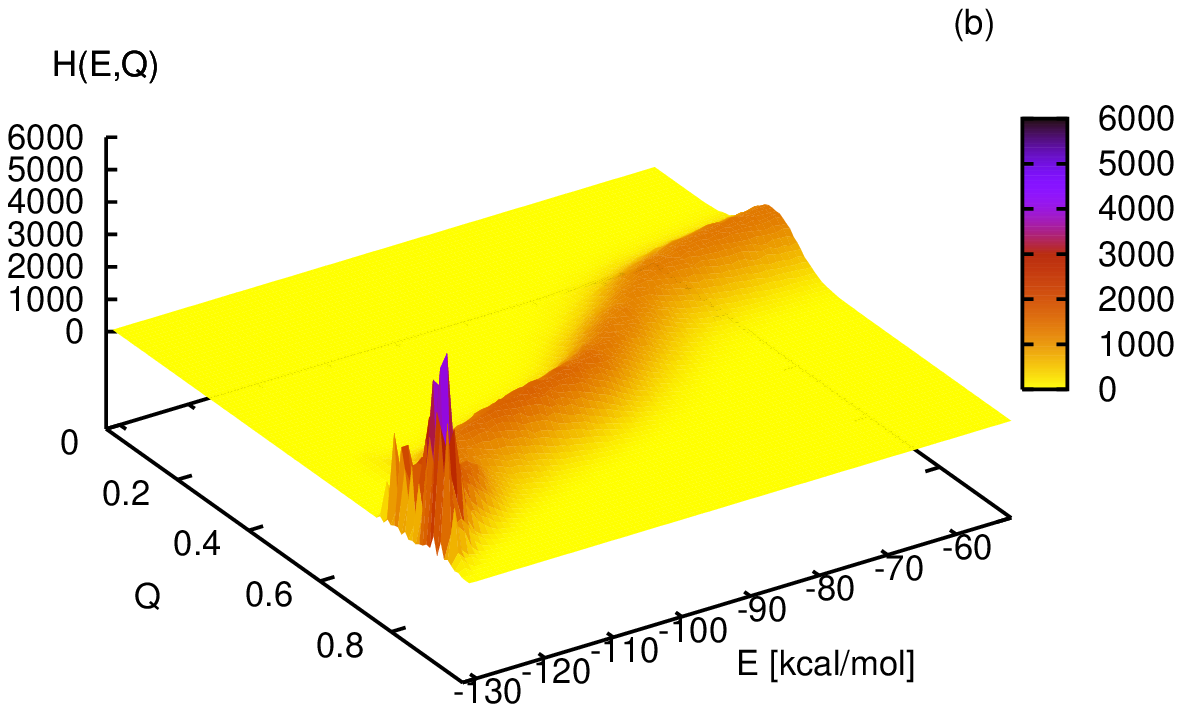}}
\centerline{\epsfxsize=8.6cm \epsfbox{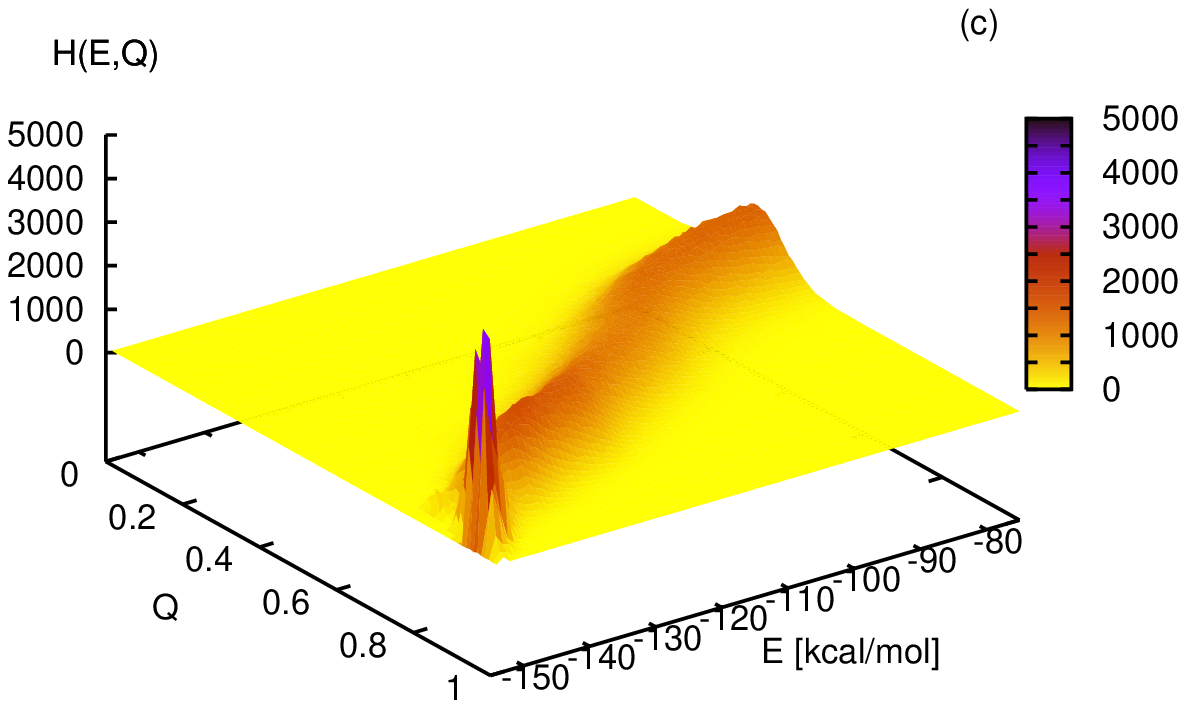}}
\caption{\label{fig:histog} (Color online)
Multicanonical histograms of overlap parameter $Q$ and energy $E$ for the peptides (a) S1, (b) S2, 
and (c) S3 in solvent.}
\end{figure}
For the study of the helix-coil folding channels, it is useful to investigate the 
two-dimensional multicanonical
histogram of the energy $E$ and a suitable system parameter, which is chosen here to be the
angular overlap parameter $Q$, as defined in Eq.~(\ref{eqov}). As reference
conformations, required for the calculation of $Q$, we use
the lowest-energy conformations found in the multicanonical simulations. These structures are shown in 
Fig.~\ref{fig:s1ref} (right) for S1 and in Fig.~\ref{fig:s23ref} for S2 and S3. 

The multicanonical histogram is obtained from the multicanonical time series 
\begin{equation}
\label{eqmuhist}
H(E,Q)=\sum\limits_t\,\delta_{E, E(t)}\,\delta_{Q, Q(t)},
\end{equation}
where the sum runs over the Monte Carlo steps $t$. The summation over $Q$ yields the 
``flat'' multicanonical energy distribution. Since the conformational energy $E$, if replaced
by the average energy $\langle E\rangle$, is directly related to the temperature, the histogram $H(E,Q)$ contains
sufficient information for the description of the simple folding transition. In Fig.~\ref{fig:histog},
we have plotted the multicanonical histograms for the three sequences S1, S2, and S3. 
In all three cases, we find a noticeable turn in the distribution from small values of $Q$, 
which correspond to random-coil conformations, to values closer to unity, where the conformations are  
similar to the helical reference conformation. In correspondence to the interpretation in the previous
section, the transition is stronger for the wild-type and mutant sequences S1 and S2, respectively, and
rather a two-step process in the case of the randomly permuted sequence S3. The remarkable bifurcations
at very low energies are possibly indications for metastable conformations which can be considered as weakly 
disturbed reference conformations. It should be noted that the angular overlap parameter is calculated
by comparing all dihedral angles. This means that deviations in side-chain dihedral angles lead to
$Q$ values different from unity, although the backbone dihedral angles could take almost the same values.
Helical structures are mainly due to cooperativity along 
the backbone, but differences in the precise side-chain locations do not destabilize the helical structures. 
Therefore, these kinds of glassy transitions, which happen under extreme conditions (very low temperatures!) 
are interesting but not in the main focus of this work.
\begin{figure}
\centerline{\epsfxsize=8.6cm \epsfbox{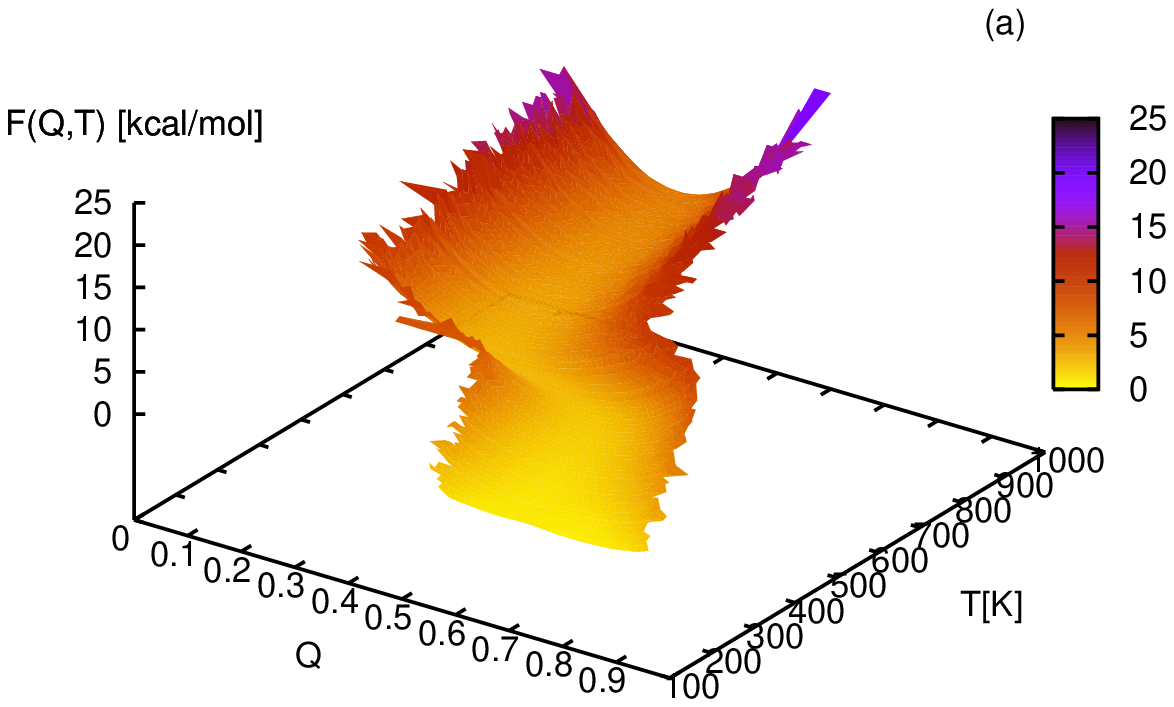}}
\centerline{\epsfxsize=8.6cm \epsfbox{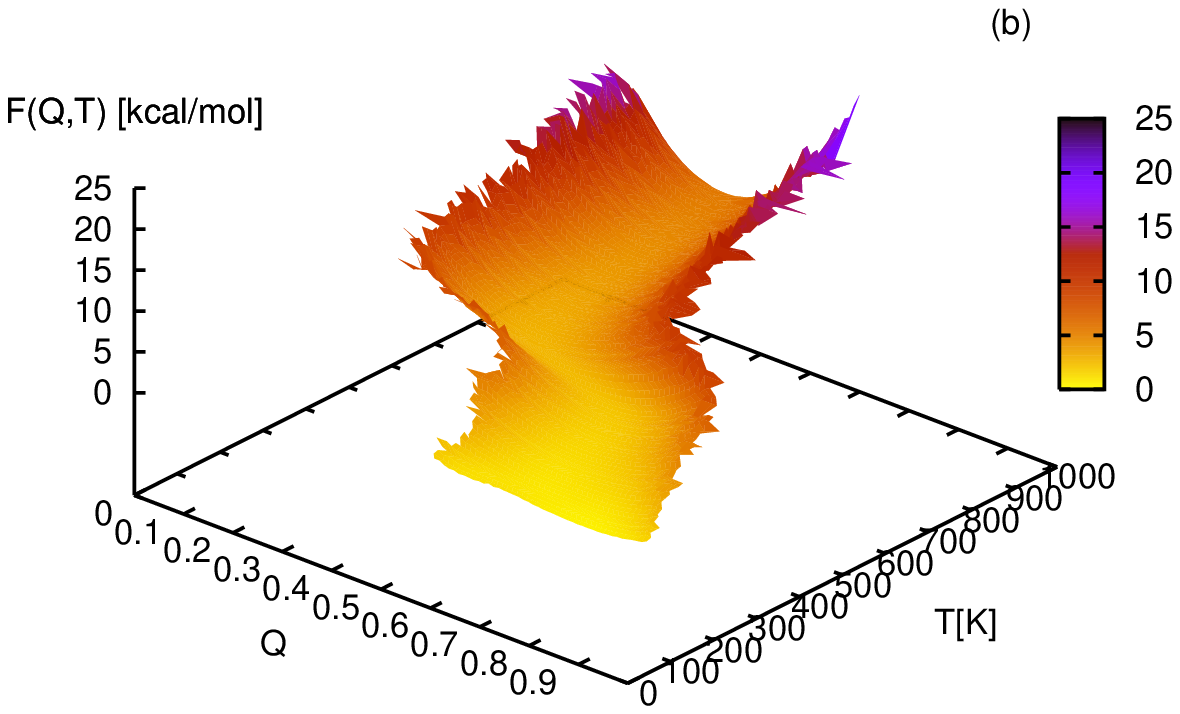}}
\centerline{\epsfxsize=8.6cm \epsfbox{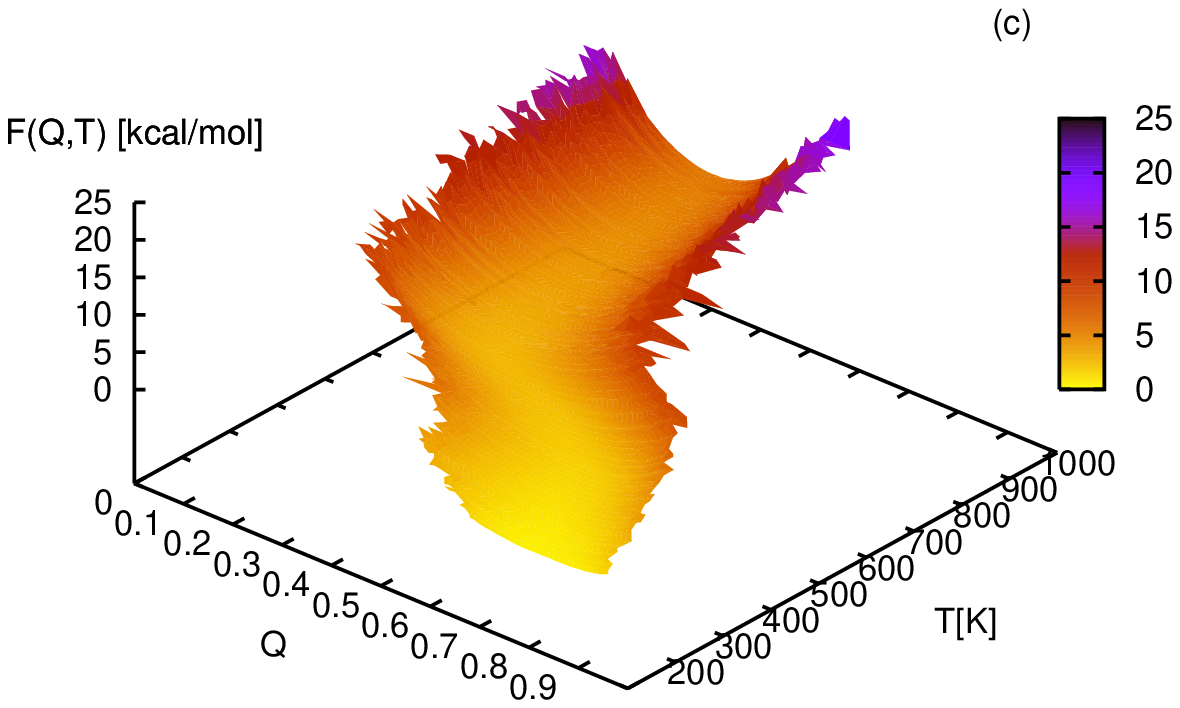}}
\caption{\label{fig:free}
(Color online) Free-energy landscapes $F(Q,T)$ for the peptides (a) S1, (b) S2, and (c) S3 in solvent.}
\end{figure}

In Figs.~\ref{fig:free}(a)--(c), we have plotted the free-energy landscapes for the three 
peptides. Here, we assume that the angular overlap parameter $Q$ is a suitable measure
for the structural order of the peptides. The free energy as a function of this ``order''
parameter and temperature is then given by 
\begin{equation}
\label{eq:free} 
F(Q,T) = -RT\ln\,p(Q,T),
\end{equation}
with the distribution of the overlap parameters 
\begin{equation}
\label{eq:pov}
p(Q,T) = \int\! {\cal D}\boldsymbol{\xi}\, \delta(Q-Q(\boldsymbol{\xi}))\,e^{-\beta E(\boldsymbol{\xi})},
\end{equation} 
where the integral runs over all possible conformations $\boldsymbol{\xi}$. The free-energy plots in 
Figs.~\ref{fig:free}(a)--(c) confirm that all three peptides experience a conformational transition
following a single main folding channel. Although the helix-coil transition separates the pseudo-phases
of random conformations and the long-range ordered helical phase, no noticeable signal of 
pseudo-phase coexistence is observed, i.e., the transition appears rather second-order- than
first-order-like. 
%
\section{Summary}
\label{secsum}
We have analysed thermodynamic properties and folding channels in the free-energy landscape for
three synthetic 12-residue peptides which exhibit remarkable adsorption affinities to
semiconductors~\cite{Goede,Whaley}. Employing an all-atom model based on the ECEPP/3 force field~\cite{ecepp}
with OONS implicit solvation parameter set~\cite{Ooi} and applying the implementation of the multicanonical 
Monte Carlo simulation method in the SMMP package~\cite{SMMP}, we found in all three cases indications for a strong
helix-coil transition. Independent of the fluctuations studied, the peak
temperatures are very close to each other -- despite of the smallness of the peptides. Since 
experimental verification and biochemical structural analysis of these peptides are still
pending, a comparison with experimental data is not yet possible. 

Our predictions for the transition temperatures are probably too high, as is to be expected by using
implicit-solvent models. Therefore, we expect that the helix-coil transitions could happen under
reasonable environmental conditions such that it should be possible to verify our predictions
experimentally. This is an important issue, since it is generally expected that selective 
synthetic peptides and polymers may play an essential
role in future nanotechnological applications.
%
\section{Acknowledgements}
We thank K.\ Goede, M.\ Grundmann, K.\ Holland-Nell, and A.\ Beck-Sickinger for interesting discussions
on peptides near substrates. We are also grateful to A.\ Irb\"ack, S.\ Mohanty, and S.\ Mitternacht
for exchanging experiences in the application of different force fields within a joint DAAD-STINT 
collaborative grant.
G.G.\ gratefully acknowledges a T\"UB\.ITAK (The Scientific \& Technological Research Council of Turkey) 
fellowship under the program-2214.
T.\c{C}.\ thanks the Alexander von Humboldt Foundation and T\"UBA for their support
and the Centre for Theoretical Sciences (NTZ) of the Universit\"at Leipzig for the
hospitality devoted to him during an extended visit.
This work is partially supported by the DFG (German Science Foundation) grant
under contract No.\ JA 483/24-1. Some simulations were performed on the
supercomputer JUMP of the John von Neumann Institute for Computing (NIC), Forschungszentrum
J\"ulich, under grant No.\ hlz11.
\end{document}